\documentclass[%
 reprint,
 amsmath,amssymb,
 aps,
]{revtex4-1}

\usepackage{graphicx}
\usepackage{dcolumn}
\usepackage{bm}
\usepackage{float}
\usepackage{footnote}

\begin{document}


\title{Structural distortions in monolayers of binary semiconductors}
\author{Poonam Kumari}
\author{Saikat Debnath}
\author{Priya Mahadevan}

\affiliation{%
 S. N. Bose National Center for Basic Sciences, Block-JD, Salt lake, Kolkata-700098\\
 }%

\date{\today}

\begin{abstract}
We examine the structural properties of free standing II-VI and III-V semiconductors at the monolayer limit within first principle density functional theory calculations. A non-polar buckled structure was found to be favoured over a polar buckled structure. While an obvious reason for this may be traced to
the contribution from dipole dipole interactions present in the 
polar structure which would  destabilize it with respect to the nonpolar structure, 
Coulomb interactions between electrons on the cations and anions are
found to be the reason for the nonpolar structure to be favoured.
A route to tune the Coulomb interaction between the electrons on the 
cations and anions is
through biaxial tensile strain. This allows for a planar graphitic 
phase in CdS to be stabilized at just 2\% tensile strain. 
Strain also shifts the valence band maximum from the $\Gamma$ point to 
the K point opening up opportunities for exploring spin-valley 
physics in these materials.
\end{abstract}

\pacs{Valid PACS appear here}
\maketitle


Polar surfaces of oxides have received a lot of interest in recent times, spurred by the good quality of oxide thin films which allowed unusual physics in this regime to be probed\cite{zno-n1,zno-n2,hwang1,hwang2,hwang3}. The field started thirty years ago with the observation of imperfect interfaces for lattice matched heterovalent semiconductors\cite{har1}. More recently with the ability to isolate graphene\citep{graphene}, the field of layered semiconductors has been rejuvenated\cite{mos21,mos22}. The inability to use graphene which is a zero band gap semiconductor in devices as we know them, has pushed the focus towards other layered semiconductors such as MoS$_2$, MoSe$_2$, WS$_2$ etc\cite{mos23,mos24}.

Binary semiconductors like ZnO have also been investigated for the possibility of having a stable 2D structure\cite{zno1}.These ionic semiconductors favor the wurtzite structure in the bulk. When films of these materials are grown without any bias for the growth direction, the polar c-direction was found to be favored. This was a surprise as one would associate a higher energy with a polar surface. On closer analysis, a graphitic phase was found to be formed\cite{zno2}, which solves the problem associated with the polar surface\cite{zno3}. In a graphitic phase one has a reduced coordination of atoms compared to the otherwise observed wurtzite phase. The system compensates for the lost coordination by decreasing the cation-anion bondlengths.

The natural question that followed was what happens to the compounds formed by elements beyond the first row at a few monolayers limit? Can the nearest neighbor bondlengths associated with these atoms with more extended wavefunctions be decreased so that a graphitic phase is realized? This question has been addressed earlier in the literature\cite{bp1,bp2}. The stability of the planar graphitic structure for the binary semiconductors involving elements beyond the first row were analyzed by studying their phonon dispersions which shows soft phonon modes indicating that these planar structures are unstable at the monolayer limit. The reason for the instability is because we cannot bring the cations and anions very close as it would increase the Coulomb repulsion between the electrons on the cations and the anions. The structure reconstructs by going to a buckled one, with the anions moving out of the plane formed by the cations. This has been pointed out earlier in the literature and a polar buckled structure has been proposed. This is surprising, as a polar surface should be energetically unfavorable. This led us to examine the phonon dispersions and we found that the deepest phonon instabilities were at M and K point. This suggested that a larger unit cell needed to be considered. A non-polar buckled structure was found to have the lowest energy. One would expect that a dominant contribution to the energy favoring the non-polar buckled structure to be associated with the absence of dipoles. Calculating the dipole moment per unit cell and evaluating this contribution to the total energy one finds that it is negligible as the electronic polarization cancels the ionic polarization. It is purely electronic considerations of a smaller Coulomb repulsion between electrons on cations and anions in the buckled non-polar structure that leads to a lower energy.

 The Coulomb repulsions can be decreased by increasing the cation-anion distance. One of the methods to do this is by application of strain. So the next question we asked was, can strain stabilize the planar graphitic structure in these materials. We took CdS as an example and it was found that a strain of 2\% could stabilize a planar graphitic structure. Strain showed some unusual effects on the electronic structure of planar CdS. A strain of 3\% was found to bring the valence band maximum (VBM) from $\Gamma$ in the unstrained case to the K-point in the strained case. This shifting of VBM from $\Gamma$ to K gives an opportunity  to explore spin-valley physics in these materials\cite{Picozzi}. The valence band maximum at K is split by spin-orbit interactions. As time reversal symmetry cannot be broken by spin-orbit interactions, the splitting at K$^{\prime}$ is opposite to that of K. This introduces an additional label to identify levels and the unusual spin-valley physics discussed in the context of transition metal dichalcogenides\cite{tmdsp} can be explored here also. This opens up an entire dimension of research in these materials.

Monolayers of III-V and II-VI semiconductors have been generated by
truncating two monolayers of cation and anion cut out from a bulk wurtzite\cite{bulk} unit cell growing in the (0001) direction.The in-plane lattice constants (ab plane) were obtained after optimization and a vacuum of 20 \AA{} has been introduced in the c-direction between images in the periodic unit cells used in our calculations.
This is needed to break the periodicity along the growth direction and thus to discard the interactions between images otherwise present as we use periodic unit cells. In order to search for possible buckled structures in the monolayer limit, an analysis of the phonon dispersions suggests a larger 6x2x1 supercell. Ground state energies have been calculated within a plane-wave implementation of density functional theory using projector augmented wave (PAW)\cite{paw1,paw2} potentials as implemented in Vienna Ab-initio Simulation Package (VASP)\cite{vasp}. We have used the local density approximation (LDA)\cite{lda}for the exchange correlation functional because earlier work in literature using LDA have found very good agreement between structural and elastic properties \cite{lda-ref1,lda-ref2}.Full geometrical optimization of internal coordinates have been done in absence of any symmetry till an energy convergence of $10^{-5}$ eV and force convergence of 5 meV/A have been achieved. A dense gamma centred mesh of 16x16x1 k-points has been used. A cutoff energy of 500 eV has been used for the plane wave basis.We have also used Hybrid functional HSE06 \cite{HSE06} with the Hartree Fock mixing factor equal to 0.2. A k-point mesh of 8x8x1 was used for these calculations. Dynamical stability of all the derived structures has been investigated by calculating the phonon spectra. Finite displacement method as implemented in the phonopy code\cite{phonopy} has been used to calculate the force constants and therefore the phonon dispersions.Supercells of 4x4x1 unit cells of the planar structure with gamma centered mesh of 8x8x1 k-points have been used to calculate force constants accurately. The acoustic sum rule has been imposed.To calculate the polarization, we have used the Berry phase method\cite{berry} as implemented in VASP

At the outset we took two monolayers of CdS (one layer each of Cd and S)cut out from a wurtzite cell to examine if a graphitic phase was stabilized. Allowing the atoms to relax and lower their energy, we found a barrierless transformation into the graphitic phase. It was found to be lower in energy than the starting wurtzite derived structure by 197 meV/formula unit (f.u). The nature of bonding in the wurtzite phase is sp$^3$ type while it is sp$^2$ type in the planar graphitic structure. Part of the energy gain in this transformation from a sp$^3$ bonded structure to a sp$^2$ bonded one comes from shortening of Cd-S bonds. The reduction in bondlengths compensates for lost coordination. These are found to be reduced by 6 \% from the wurtzite phase. We then went on to probe if this planar graphitic structure was stable.To study its stability, the phonon dispersions were calculated. Fig.1 shows the phonon dispersions for planar CdS. Mode softening for one of the acoustic phonon mode is found along the entire Brillouin zone. This suggests that the planar graphitic structure is not stable. This can be attributed to the fact that the spatial extension of the wavefunction of the valence orbitals of the anions is large for the compounds involving elements beyond the first row. The Coulomb interaction between the electrons on the anions and cations is therefore large. In order to reduce this repulsion, the anions try to move away from the cations by moving out of the plane and the system no longer remains planar. This led to a buckled structure being proposed in the literature\cite{bp1,bp2}, which has cations in one plane and anions in the other. As a result, the system acquires a dipole moment.
Consequently, as more layers are added, the surface energy is expected to diverge and the system tries to move away from this point.Therefore polar structures are not expected to be realized beyond a few monolayers\cite{harrison}. Various mechanisms like surface reconstructions, adsorption of adatoms,vacancy formations, transfer of charges etc. may help to overcome this polar divergence and for a non-polar structure to be stabilized\cite{noguera}. We start by examining the structures favoured at the monolayer limit.

In order to understand this, we took a closer look at the phonon dispersions. It was found that the deepest phonon instabilities are at M and K points. These have been found to arise from the force on S ion trying to move them out of plane. The non $\Gamma$ character of the deepest phonon instabilities suggest that the movement of atoms are in the opposite directions in the neighboring unit cells. This means if a S atom moves out of plane in the upward direction in one unit cell, the S atom in the adjacent unit cell will move in the downward direction. Considering this movement of anions, we constructed a supercell of 6x2x1 and allowed the atoms to move out of the plane as shown in Fig.2 Anions in the plane above and below the cation plane have been indicated with Up and Dn respectively. As the unit cell in this case was large, we calculated the phonon dispersion at the gamma point only and the phonon modes were found to be positive. On examining this buckled structure, it was found that it has no net dipole moment as opposed to the previously reported buckled structure\cite{bp1,bp2}.We considered a number of II-VI and III-V binary semiconductors and constructed similar structures for each of them. Energies per formula unit of each of these structures were calculated and are compared in Table 1. Energies per formula unit of both the structures were also calculated using hybrid functionals (HSE06).It can be seen that the non-polar buckled structure (BNP) is indeed lower in energy than the polar buckled structure (BP) in all the cases, indicating that even at the monolayer limit these semiconductors favor a non-polar structure. One might associate surface energy divergence from a polar surface being the reason for the non-polar buckled structure being lower in energy with respect to buckled polar structure. In order to investigate this, we calculated the dipole moment associated with  the polar buckled structure. On analysis we found that indeed the ionic dipole moment is high but the electronic polarization almost compensates the ionic dipole moment. Hence the net dipole moment is very small and is not enough to sustain an electric field. We calculated the net dipole moment and the energy due to this (Table 2). It can be seen that the energy due to the dipole moments are very small. To understand the reason behind the stability of the non-polar buckled structure we examined the bondlengths in both the cases. The bondlengths are listed in Table 3. It can be seen that in the non-polar buckled structure, the cation-anion as well as the anion-anion bondlengths have increased compared to the polar buckled structure, due to which the Coulomb repulsion between the electrons on different atoms reduces, which lowers the energy of the non-polar buckled structure\cite{energy-suppl}. In order to further verify if that is the case, we constructed non-polar buckled structure keeping the cation-anion bondlenths same as that of the polar buckled structure and calculated the energies. It was seen that the difference in energies are now very small and can be attributed to the difference in anion-anion bondlenths. Hence it can be concluded from this analysis that the only reason for the non-polar buckled structure to be favored is the increase in bondlenth which reduces the Coulomb repulsion between the electrons on different atoms.

Another method of decreasing the Coulomb repulsion between the electrons is by subjecting the system to biaxial tensile strain. A strain of 2\%, 3\% and 4\% was applied on planar CdS monolayer and their stabilities were checked by studying their phonon dispersions. Fig.3 shows phonon dispersions for 2 \% strained planar graphitic structure. It can be seen that only positive phonon modes are present, indicating that the structure is stable. Thus we have a rare realization of a sp$^2$ bonded phase for CdS. We then went on to examine the effect of biaxial tensile strain on the electronic structure of CdS monolayer. The unstrained films are found to have their valence band maximum at $\Gamma$. The highest occupied band at K is $\sim$ 0.2 eV deeper inside the valence band (Fig.4). An analysis of the charge density (Fig.5) reveals that the valence band maximum emerges from interactions between the Cd d$_{x^2-y^2}$,d$_{xy}$ and S p$_y$ orbitals. However the highest occupied band at K is contributed by anion-anion interactions between the S p$_z$ orbitals. An empirical scaling law given by Harrison provides a relation of how the hopping interaction strengths would scale with distance. According to this law, the $pd$ interactions scale as $\dfrac{1}{r^{4}}$ while the $pp$ interactions scale as $\dfrac{1}{r^{3}}$. This immediately provides us with a handle of being able to control this separation. Introducing an in-plane tensile strain, we find that, a 3\% tensile strain pushes the highest occupied band at $\Gamma$, $\sim$ 0.16 eV deeper. This emerges from the fact that these states are from antibonding Cd d$_{x^2-y^2}$,d$_{xy}$ and S p$_y$ interactions which moves deeper into the valence band as the strength of this interaction is decreased. The variations are depicted pictorially in Fig. 4, where the electrostatic potential on a Cd atom are used as an internal energy reference to compare the movement of the highest occupied band between different calculations. There seems to be an apparent movement of the highest occupied band at K to higher energies. The movements are much smaller than that of the band at $\Gamma$, suggesting that other effects such as charge transfer between Cd and S could modify the onsite energies, and lead to the observed movement of the band to higher energies. The important consequence of applying strain is that the band extrema shifts to K point at 3\% strain. As discussed in the introduction, spin-orbit effects lead to a spin splitting of the highest occupied band at K point. This is found to be $\sim$20 meV in CdS and increases to $\sim$50 meV in CdTe. This would then make these systems also important candidates for probing spin valley physics, and additionally introduce drastic change in their transport behavior. A substrate could easily provide the required tensile strain and future work will examine ultrathin films of these semiconductors on appropriate substrates.

In conclusion, the present investigation examines the structure favoured by monolayers of group II-VI and III-V semiconductors. A non-polar buckled structure has been found to be energetically more favourable than a polar buckled one. Reduced Coulomb repulsion between the electrons on the anion and cations because of increased bondlengths in the non-polar structure compared to the polar one plays a major role in stabilizing it. A biaxial tensile strain of 2\% is capable of eliminating the buckling in such structures stabilizing a graphitic analogue. Additionally, with 3\% biaxial tensile strain, the position of the VBM shifts from $\Gamma$ to K point which makes them suitable candidates for exploring spin valley physics.
 
\section{\label{sec:level1}ACKNOWLEDGEMENT}
P.M. thanks DST Nanomission for support
through a project as well as for computational resources under
the umbrella of the Thematic Unit of Computational Material
Science at the S. N. Bose centre.

PK and SD have contributed equally to this work.


\begin{figure}[H]
\includegraphics[width=3 in]{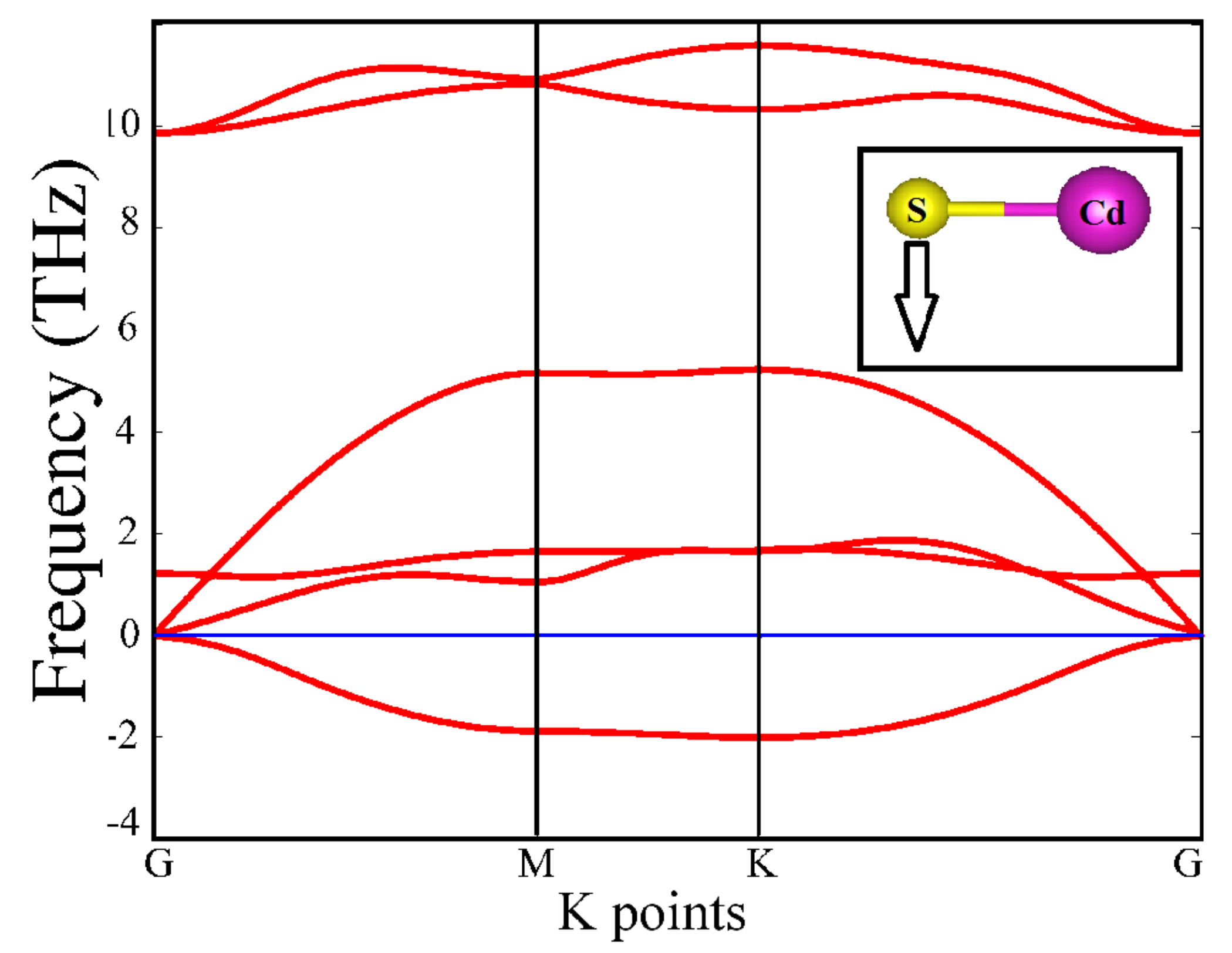}
\caption{The phonon dispersion of graphitic phase of a monolayer of CdS along different symmetry directions using optimized in-plane lattice constants. The displacement of the atoms corresponding to the eigenfunction for the unstable phonon mode in one unit cell are shown in the inset }
\end{figure}

\begin{figure}[H]
\includegraphics[width=3 in]{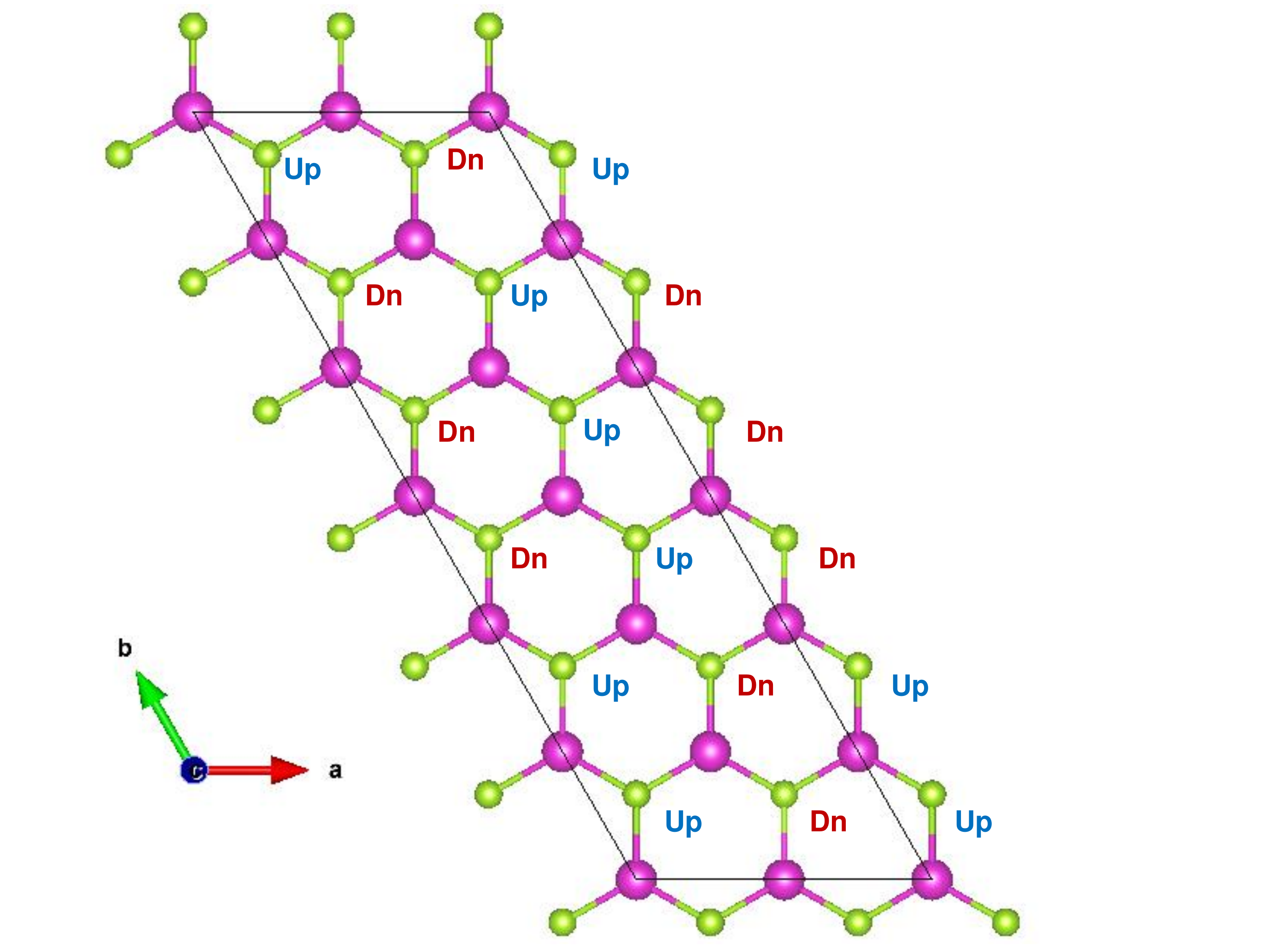}
\caption{Top view of relaxed buckled non-polar structure for CdS where Up(Dn) denotes the anions that have moved in positive(negative) z-direction.}
\end{figure}

\begin{table}[H]
\centering
\begin{tabular}{|c|c|c|c|}
\hline
•& E(BP-P)
eV
 & E(NBP-BP)
eV
 & E(NBP-BP)
eV (HSE06)
 \\
\hline
CdS & 0.000 & -0.015 & -0.016 \\
\hline
CdSe  & -0.008 & -0.049 & -0.060 \\
\hline
CdTe  & -0.027 & -0.062 & -0.083 \\
\hline
ZnS & 0.000 & -0.011 & -0.032\\
\hline
ZnSe & -0.001 & -0.039 & -0.054\\
\hline
ZnTe & -0.014 & -0.050 & -0.070\\
\hline
AlP & 0.000 & -0.018 & -0.007\\
\hline
AlAs & -0.016 & -0.030 & -0.034 \\
\hline
GaP & -0.020 & -0.029 & -0.037\\
\hline
GaAs & -0.091 & -0.025 & -0.042 \\
\hline
\end{tabular}
\caption{Table comparing the energies of the Planar(P), Buckled Polar (BP) and Buckled Non-Polar(BNP) structures for a monolayer of the listed semiconductor.}
\end{table}

\begin{table}[H]
\begin{tabular}{|c|c|c|}
\hline 
• & Net Dipole moment e\AA & E(dp)meV \\ 
\hline 
CdS & 0.0003 & 0.0 \\ 
\hline 
CdSe & 0.0824 & 1.2 \\ 
\hline 
CdTe & 0.0943 & 1.3 \\ 
\hline 
ZnS & 0.0004 & 0.0 \\ 
\hline 
ZnSe & 0.0409 & 0.3 \\ 
\hline 
ZnTe & 0.0674 & 0.8 \\ 
\hline 
AlP & 0.0007 & 0.0 \\ 
\hline 
AlAs & 0.0832 & 1.5 \\ 
\hline 
GaP & 0.0712 & 1.2 \\ 
\hline 
GaAs & 0.0805 & 1.4 \\ 
\hline 
\end{tabular}
\caption{Table showing the net dipole moment and the energy due to the dipole (E(dp)) for the buckled polar structure .}
\end{table}

\begin{table}[H]
\centering
\begin{tabular}{|c|c|c|c|c|}
\hline
• &\multicolumn{2}{c|}{Polar Buckled Structure}& \multicolumn{2}{c|}{Non-Polar Buckled Structure} \\
\hline
• & Cation-Anion & Anion-Anion & Cation-Anion & Anion-Anion \\
\hline
CdS & 2.40 & 4.16 & 2.44 & 4.23 \\
\hline
CdSe & 2.51 & 4.32 & 2.54 & 4.47 \\
\hline
CdTe & 2.70 & 4.59 & 2.74 & 4.79 \\
\hline
ZnS & 2.20 & 3.80 & 2.22 & 3.85 \\
\hline
ZnSe & 2.31 & 3.99 & 2.36 & 4.09 \\
\hline
ZnTe & 2.50 & 4.29 & 2.55 & 4.44 \\
\hline
AlP & 2.24 & 3.89 & 2.28 & 3.96 \\
\hline
AlAs & 2.35 & 4.03 & 2.39 & 4.16 \\
\hline
GaP & 2.28 & 3.87 & 2.29 & 3.99 \\
\hline
GaAs & 2.38 & 4.01 & 2.40 & 4.20 \\
\hline
\end{tabular}
\caption{Table showing the cation-anion and anion-anion bondlengths (in \AA) for both polar buckled and non-polar buckled structures}
\end{table}

\begin{figure}[H]
\includegraphics[width=3 in]{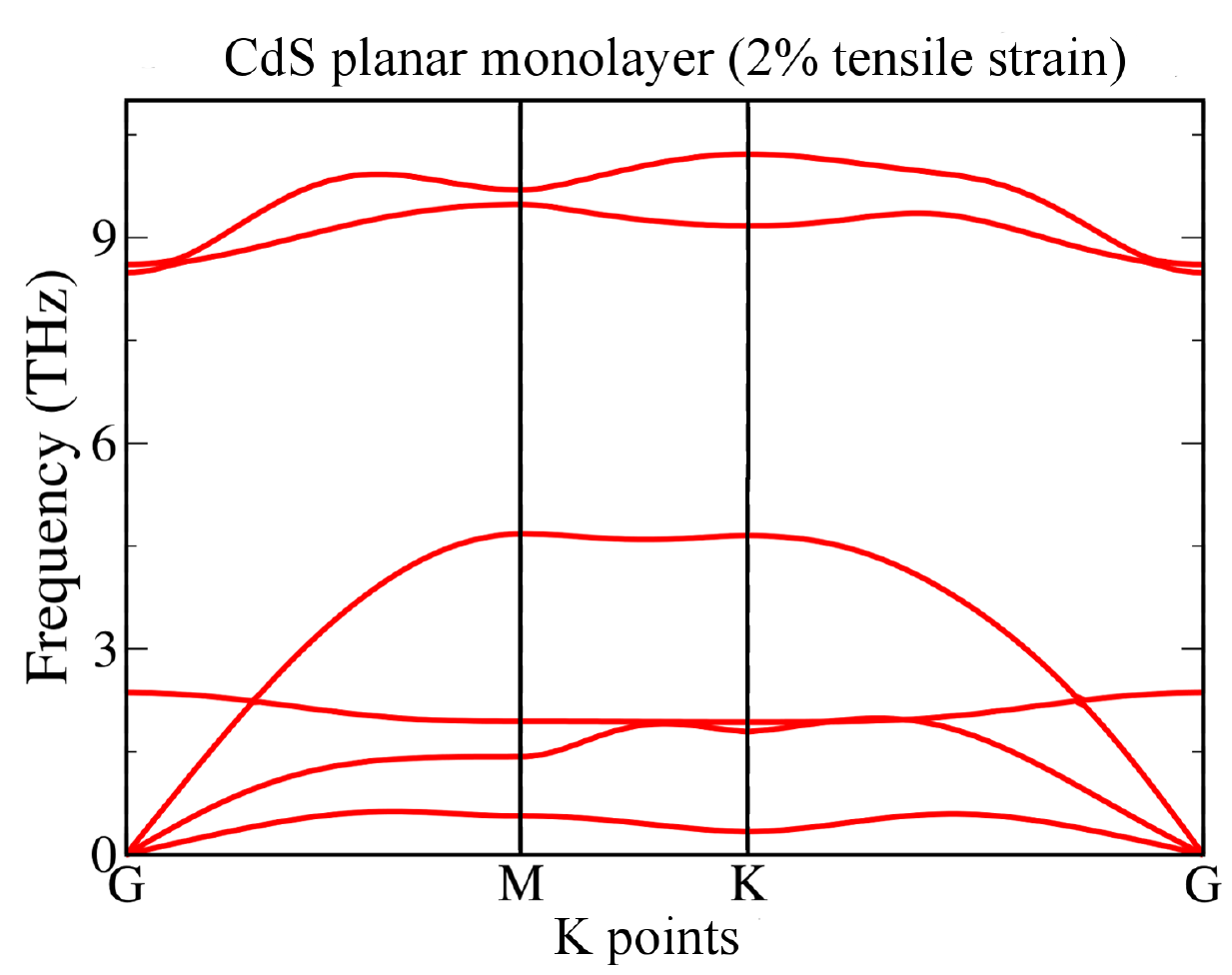}
\caption{Phonon dispersion of monolayer of planar CdS under 2\% biaxial tensile strain.}
\end{figure}

\begin{figure}[H]
\includegraphics[width=3 in]{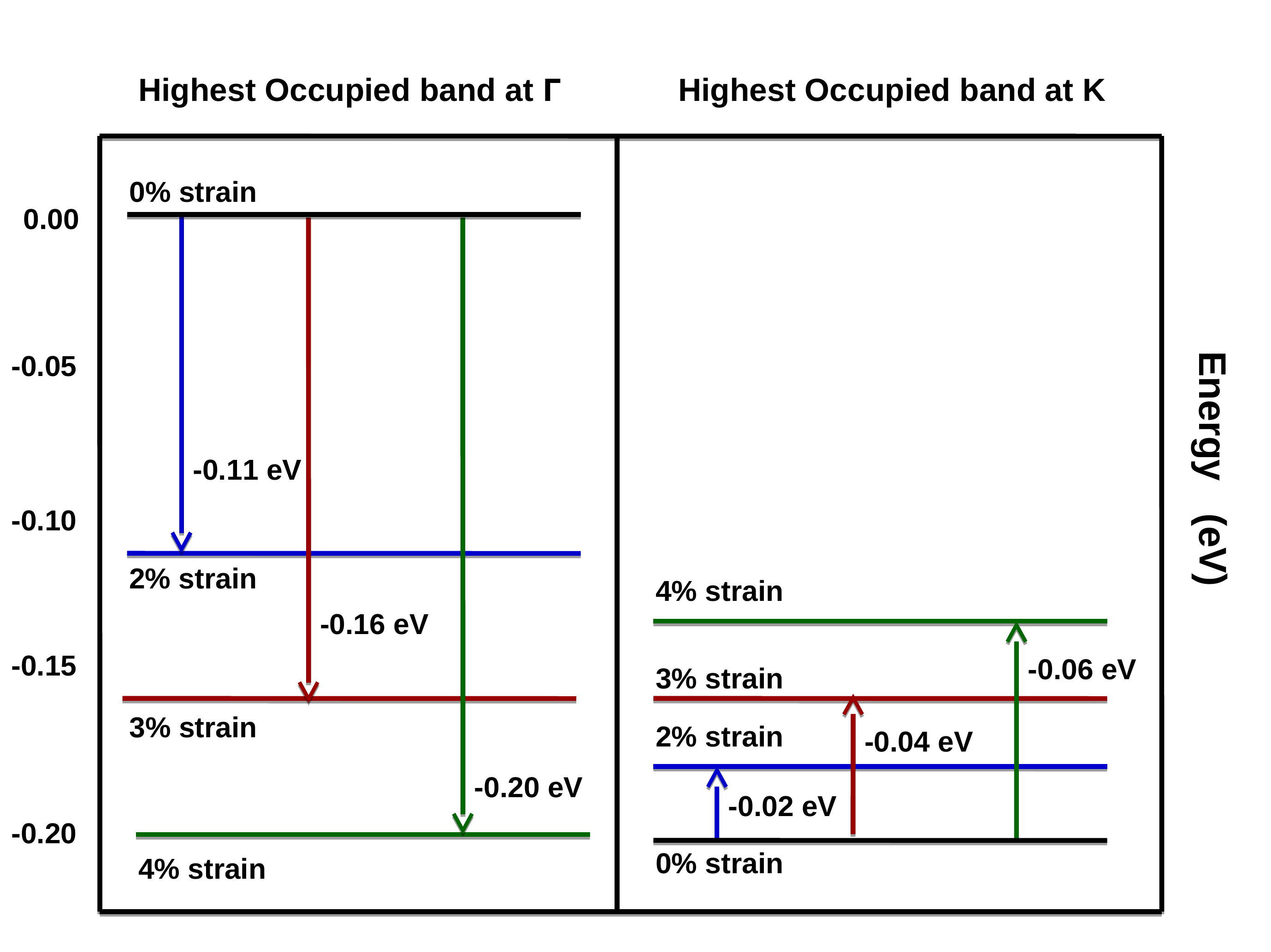}
\caption{Shift in Energy (in eV) of the highest occupied bands at $\Gamma$ and K on application of strain in planar CdS.}
\end{figure}

\begin{figure}[H]
\includegraphics[width=3 in]{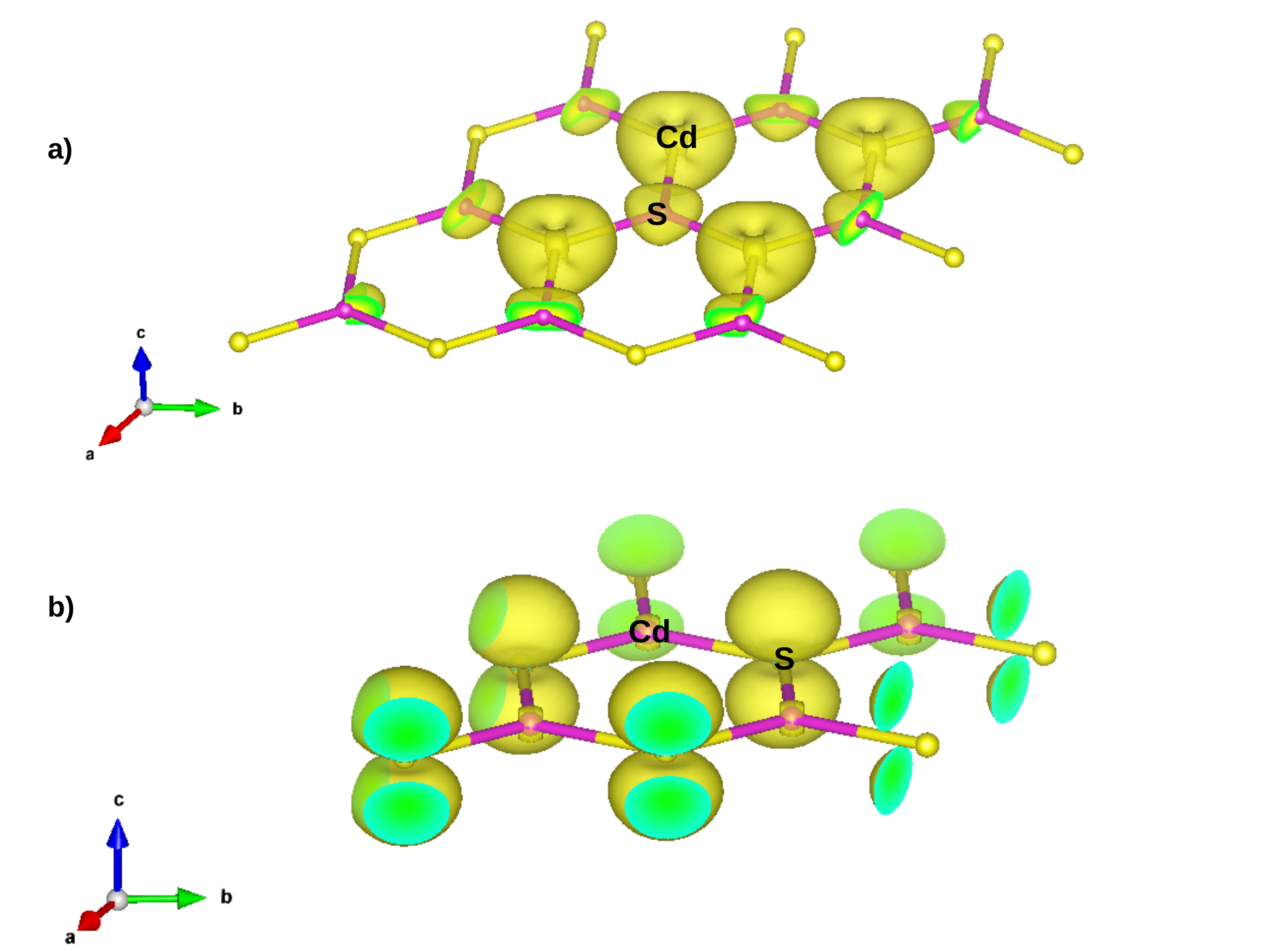}
\caption{a)Charge density of the highest occupied band at $\Gamma$ b) Charge density of the highest occupied band at K-point in  planar CdS}
\end{figure}


\begin{thebibliography}{20}

\bibitem{zno-n1}
C.Noguera and G.Goniakowski,  J. Phys.: Condens. Matter \textbf{20}, 264003 (2008).

\bibitem{zno-n2}
G.Goniakowski, F.Finocchi and C.Noguera  Rep. Prog. Phys. \textbf{71}, 016501 (2008).

\bibitem{hwang1}
A. Ohtomo, D. A. Muller, J. L. Grazul, and H. Y. Hwang,  Nature \textbf{419}, 378 (2002).

\bibitem{hwang2}
H. Y. Hwang, Y. Iwasa, M. Kawasaki, B. Keimer, N. Nagaosa, and Y. Tokura,  Nature Mat. \textbf{11}, 103 (2012).

\bibitem{hwang3}
N. Nakagawa, H. Y. Hwang, and D. A. Muller,  Nature Mat. \textbf{5}, 204 (2006).

\bibitem{har1}
W. A. Harrison, E. A. Kraut, J. R. Waldrop, and R. W. Grant, Phys. Rev. B \textbf{18}, 4402 (1978).

\bibitem{graphene}
K. S. Novoselov, A. K. Geim, S. V. Morozov, D. Jiang,Y. Zhang,S. V. Dubonos, I. V. Grigorieva, and A. A. Firsov, Science \textbf{306}, 666 (2004).

\bibitem{mos21}
B. Radisavljevic, A. Radenovic, J. Brivio, V. Giacometti, and A. Kis, Nat. Nanotechnol. \textbf{6}, 147 (2011).

\bibitem{mos22}
Y. Zhang, J. Ye, Y. Matsuhashi, and Y. Iwasa, Nano Lett. \textbf{12}, 1136 (2012).

\bibitem{mos23}
Z. Yin, Hai Li, Hong Li, L. Jiang, Y. Shi, Y. Sun, G. Lu, Q. Zhang, X. Chen, and H. Zhang, ACS Nano \textbf{6}, 74 (2012).

\bibitem{mos24}
H. Fang, S. Chuang, T. C. Chang, K. Takei, T. Takahashi, and A. Javey, Nano Lett. \textbf{12}, 3788 (2012).

\bibitem{zno1}
F. Claeyssens, C. L. Freeman, N. Allan, Y. Sun, M. N. R. Ashfolda, and J. H. Harding, Journal of Mater. Chem., \textbf{15}, 139 (2005).

\bibitem{zno2}
C. Tusche, H. L. Meyerheim, and J. Kirschner, Phys. Rev. Lett. \textbf{99},  026102 (2007).

\bibitem{zno3}
B. Rakshit, and P. Mahadevan, Phys. Rev. Lett. \textbf{107 },  085508 (2011).

\bibitem{bp1}
Houlong L. Zhuang, Arunima K. Singh, and Richard G. Hennig Phys. Rev. B. \textbf{87},  165415 (2013).

\bibitem{bp2}
Hui Zheng, Xian-Bin Li, Nian-Ke Chen, Sheng-Yi Xie, Wei Quan Tian, Yuanping Chen, Hong Xia, S. B. Zhang,  and Hong-Bo Sun, Phys. Rev. B. \textbf{92},  115307 (2015).

\bibitem{Picozzi}
Domenico Di Sante, Alessandro Stroppa, Paolo Barone, Myung-Hwan Whangbo, and Silvia Picozzi, Phys. Rev. B. \textbf{91},  161401 (2015).

\bibitem{tmdsp}
Z. Y. Zhu and Y. C. Cheng and U. Schwingenschlögl, Phys. Rev. B. \textbf{84},  153402 (2011).

\bibitem{freeman}
Colin L. Freeman, Frederik Claeyssens, Neil L. Allan and John H. Harding, Phys. Rev. Lett. \textbf{96}, 066102 (2006).

\bibitem{bulk}
Y. Liu, Y. Xu, J.P. Li, B. Zhang, D. Wu and Y.H. Sun, Mater. Res. Bull. \textbf{41}, 99 (2006).

\bibitem{paw1}
G. Kresse and D. Joubert, Phys. Rev. B \textbf{59}, 1758 (1999).

\bibitem{paw2}
P. E Bl\"{o}chl, Phys. Rev. \textbf{50}, 17953 (1994).

\bibitem{vasp}
G. Kresse and J. Furthm\"{u}ller, Phys. Rev. B \textbf{54}, 11169 (1996).

\bibitem{lda}
J. P. Perdew, and A. Zunger, Phys. Rev. B \textbf{23}, 5048 (1981).

\bibitem{lda-ref1}
Su-Huai Wei and S. B. Zhang Phys. Rev. B \textbf{62},  6944 (2000).

\bibitem{lda-ref2}
Su-Huai Wei and Alex Zunger Phys. Rev. B \textbf{60},  5404 (1999).

\bibitem{HSE06}
J. Heyd, G.E. Scuseria, and M. Ernzerhof, J. Chem. Phys.
\textbf{118} 8207 (2003); J. Chem. Phys. \textbf{124} 219906 (2006).

\bibitem{phonopy}
Atsushi Togo, Fumiyasu Oba, and Isao Tanaka, Phys. Rev. B \textbf{78}, 134106 (2008).

\bibitem{berry}
R. W. Nunes and Xavier Gonze, Phys. Rev. B \textbf{63}, 155107 (2001).

\bibitem{harrison}
W. A. Harrison, E. A. Kraut, J. R. Waldropand R. W. Grant, Phys. Rev. B \textbf{18}, 8 (1978).

\bibitem{noguera}
C. Noguera and J. Goniakowski, Chem. Rev. \textbf{113}, 6 (2012).

\bibitem{energy-suppl}
These ideas are supported by a breakup of the total energy into various components and given in the supplementary information.

\end{thebibliography}
\end{document}